\documentclass{aa}  

\usepackage{orcidlink}
\usepackage{graphicx}
\usepackage{xcolor}
\usepackage{soul}

\usepackage{tikz}
\usetikzlibrary{arrows.meta, positioning, fit}

\usepackage{txfonts}
\newcommand{\healpix}[0]{\textsc{HEALPix}}
\newcommand{\pkdgrav}[0]{\textsc{pkdgrav3}}
\usepackage{hyperref}
\hypersetup{
    colorlinks=true,
    citecolor=blue,     % citation links
    linkcolor=blue,    % internal links (sections)
    urlcolor=blue       % URLs
}

\begin{document}

   \title{Integrated Sachs–Wolfe maps from the Gower Street $w$CDM simulations}

   \author{Mina Ghodsi Yengejeh,
          \inst{1,2,3}\orcidlink{0000-0001-5481-9810}
            \and
        András Kovács, \inst{1,2}\orcidlink{0000-0002-5825-579X}
            \and
        Istv\'an Szapudi,
        \inst{4}\orcidlink{0000-0003-2274-0301}
           \and
        Istv\'an Csabai
        \inst{3}\orcidlink{0000-0001-9232-9898}
          }

    \institute{MTA--CSFK \emph{Lend\"ulet} ``Momentum'' Large-Scale Structure (LSS) Research Group, Konkoly Thege Mikl\'os \'ut 15-17, H-1121 Budapest, Hungary
    \and
    Konkoly Observatory, HUN-REN Research Centre for Astronomy and Earth Sciences, Konkoly Thege Mikl\'os \'ut 15-17, H-1121 Budapest, Hungary
    \and
    Institute of Physics and Astronomy, ELTE E\"otv\"os Lor\'and University, P\'azm\'any P\'eter s\'et\'any 1/A, H-1117 Budapest, Hungary
    \and
    Institute for Astronomy, University of Hawaii, 2680 Woodlawn Drive, Honolulu, HI 96822, USA
    }
    
   \date{Received Month Day, 2025; accepted Month Day, Year}

  \abstract
   {The late-time linear Integrated Sachs-Wolfe (ISW) effect directly probes the dynamics of cosmic acceleration and the nature of dark energy. Detecting these weak, secondary temperature anisotropy signals of the Cosmic Microwave Background requires accurate theoretical predictions of their amplitude across cosmological models.}
   {By extending the \texttt{pyGenISW} package, previously limited to $\Lambda$CDM, we aim to generate full-sky ISW maps for a suite of 791 $w$CDM cosmologies using the Gower Street N-body simulations, thereby enabling ISW analyses across a broader dark-energy parameter space. We make our code and ISW data publicly available.} 
   {We compute the ISW signals by tracing the time evolution of the gravitational potential across large-volume simulations that span dark energy equation of state parameters from phantom to quintessence, $-1.79 \lesssim w \lesssim -0.34$. These data are projected onto the sphere using \healpix{} to obtain full-sky temperature maps.}
   {We validate our pipeline by comparing the measured ISW angular power spectra and ISW-density cross-correlations against linear theory expectations ($2 \leq \ell \leq 200$) computed with benchmarks from the \texttt{pyCCL} library. The agreement is excellent across the multipole range where the ISW contribution is expected to dominate, confirming the reliability of our modelling of gravitational-potential evolution.  With additional tests of the ISW signal's strength in density extrema, as well as comparing all models to a reference $\Lambda$CDM cosmology using power ratios, we found that quintessence-like models ($w > -1$) show higher ISW amplitudes than phantom models ($w < -1$), consistent with enhanced late-time decay of gravitational potentials.}
   {The consistency of our $w$CDM ISW maps and their agreement with theory predictions confirm the robustness of our methodology, establishing it as a reliable tool for theoretical and observational ISW-LSS analyses. This includes applications to next-generation surveys in the context of covariance calculations and various map-based statistics.}


   \maketitle
%
%-------------------------------------------------------------------

%%%%%%%%%%%%%%%%%%%%%%%%%%%%
%%%%%%% Introduction %%%%%%%
%%%%%%%%%%%%%%%%%%%%%%%%%%%%
\section{Introduction}
\label{sec:intro}
The accelerated expansion of the Universe at low redshifts, and the corresponding suppression of the growth of cosmic structure due to dark energy,  are among the most intriguing mysteries in modern cosmology \citep[see e.g.,][]{RiessEtal98}. Combined with cold dark matter (CDM) in the consensus $\Lambda$CDM model, a cosmological constant $\Lambda$, i.e., the simplest model for dark energy, fits an impressive range of observations at various time and length scales. Yet, the nature of these dark substances remains unknown, which motivates further research on both theoretical and observational fronts \citep[see e.g.,][]{Abdul_Karim_2025}.

Given the conceptual challenges posed by $\Lambda$, i.e., fine-tuning \citep{Weinberg:1988cp}, the cosmic coincidence problem \citep{75796370-3c29-39fe-a27e-514cd98f9469}, and the wide range of theoretical alternatives (i.e., dynamical dark energy, interacting dark sectors, modified gravity) \citep[see e.g.,][for review]{odintsov2024}, it is essential to probe cosmic acceleration from every possible angle and with as many independent data sets as possible. Different cosmological probes are sensitive to different physical aspects of dark energy, from background expansion to the large-scale clustering of matter. A coherent picture may only emerge when these perspectives are combined and cross-checked, in line with the strategies of several current and upcoming cosmological sky survey projects \citep[see e.g.,][]{DESI,euclid}.

One particular probe of dark energy via the delicate, redshift-dependent balance of structure growth and cosmic expansion is the Integrated Sachs-Wolfe (ISW) effect \citep{Sachs1967}. On top of the primary temperature anisotropies from the recombination era, the ISW signal is a secondary anisotropy of the Cosmic Microwave Background (CMB) radiation, offering a unique window on the dynamical properties of cosmic acceleration. On their journey towards our telescopes, CMB photons traverse galaxy superclusters and voids in the cosmic web \citep[see e.g.,][]{NadathurCrittenden2016, Kovacs2022}. They experience blueshifts when falling into a potential well of a supercluster and redshifts when exiting it. In contrast, photons experience redshifts when climbing on a potential hill of a void and blueshifts during their descent. If the depth of the potential does not change with time, like in a matter-dominated Einstein-de Sitter Universe, these blueshifts and redshifts cancel out at the linear level \citep[see also][for non-linear effects]{Rees1968}. 

However, if the large-scale gravitational potential decays or grows, photons will end up with a \emph{net} energy shift, resulting in additional CMB temperature perturbations via the ISW effect, generated by the large-scale matter fluctuations at about $\sim100 \, h^{ - 1} \mathrm{Mpc}$ scales. This shift arises because the ISW signal accumulates along the line-of-sight (LOS), corresponding to an integral over the time evolution of the gravitational potential from the last scattering surface to today. When transitioning from matter domination to dark energy domination at approximately $z<1.5$, the late-time ISW effect is sourced by the extra space-stretching effects of dark energy, and its detection is of great interest in modern cosmology.

To detect the tiny ISW signal at the $\Delta T_{\rm ISW}\sim 1-10\,\mu K$ level, one must cross-correlate the large-angle CMB anisotropies with maps of the cosmic large-scale structure. In order to quantify the results, it is common to introduce an amplitude parameter, $A_{\rm ISW}$, defined operationally in cross-correlation analyses as the scaling between the measured ISW strength and the prediction from a fiducial $\Lambda$CDM model
\begin{equation}
    A_{\rm ISW} := \dfrac{\text{Measured ISW signal from data}}{\text{Predicted ISW signal in fiducial $\Lambda$CDM}}.
    \label{eq:A_ISW}
\end{equation}
If the fiducial $\Lambda$CDM model were correct, then we would expect $A_{\rm ISW} \approx 1$, meaning that the observed ISW signal matches the predicted imprint from the decay of gravitational potentials in a $\Lambda$-dominated era. Values of $A_{\rm ISW}$ that differ from unity indicate one of the following: statistical fluctuations, unaccounted observational or modeling systematics, deviation from $\Lambda$CDM model in terms of expansion and/or growth history \citep[see e.g.,][and references therein, for ISW modeling in alternative models]{Schafer_2008,CaiEtAl2014,Velten:2015qua,Carbone2016,Mostaghel:2018pia,Adamek2020,Ghodsi_Yengejeh_2023}. 

Traditional detection methods include 2-point correlation measurements and cross-power spectra, using larger and larger galaxy catalogs over the last two decades \citep[see e.g.,][and references therein]{cooray, Fosalba2003, cabre_etal, gianEtal08,Xia:2009dr,GranettEtal2009, Dupe_2011,rassat2013,PlanckISW2015,Stolzner2018, Ansarinejad_2020, BahrKalus_2022}. The majority of these results are consistent with $A_{\rm ISW}\approx 1$ and thus with the basic $\Lambda$CDM model, albeit presenting low-to-moderate detection significances. As a state-of-the-art result, a recent ISW analysis by \cite{Hang20212pt} reported $A_{\rm ISW}\approx 0.984\pm0.349$, corresponding to a $2.8\,\sigma$ detection, using the Dark Energy Spectroscopic Instrument (DESI) Legacy Survey galaxy catalogs. This finding is again fully consistent with $\Lambda$CDM expectations.

Since the ISW signals are intrinsically weak, a complementary approach is \emph{stacking} cutout images of the CMB temperature map in the positions of specific large-scale structures, such as voids and superclusters, to focus on the most extreme parts of the cosmic web where most of the ISW imprints are generated. 

A series of such stacking analyses on super-structures detected hot (cold) CMB imprints associated with superclusters (voids) identified in galaxy catalogs, pioneered above all by \cite{Granett2008} who reported a $\sim 4.4\,\sigma$ detection using Sloan Digital Sky Survey (SDSS) data and Wilkinson Microwave Anisotropy Probe (WMAP) CMB data. This was later confirmed using \emph{Planck} observations too \citep{PlanckISW2015}. 

These findings were especially intriguing, because theoretical analyses have concluded that the observational ISW signal from super-structures is significantly higher than expected in the basic $\Lambda$CDM model \citep[see e.g.,][]{Nadathur2012,Hernandez2013,Ilic2013}. Several further studies have revisited and extended this approach:

\begin{itemize}
    \item[\textbullet] Using galaxy catalogs from the Baryon Oscillations Spectroscopic Survey (BOSS) and the Dark Energy Survey (DES), some of the results again reported excess ISW signal from the largest voids in particular \citep[see e.g.,][]{CaiEtAl2013,Kovacs2016,Kovacs2018}, confirming the findings by \cite{Granett2008} in other parts of the sky and using alternative void catalogs.
    \smallskip
    \item[\textbullet] In particular, \cite{Kovacs2019} found $A_{\rm ISW}\approx 5.2\pm1.6$ ($3.3\,\sigma$ significance) using $R\geq100\, h^{ - 1} \mathrm{Mpc}$ supervoids, identified in the BOSS Data Release 12 and DES Year-3 data sets, and combined into a joint measurement. 
    \smallskip
    \item[\textbullet] Using alternative void finding and stacking techniques applied to the BOSS galaxy catalogs, however, \cite{NadathurCrittenden2016} reported no significant tension with $A_{\rm ISW}\approx 1.64\pm0.53$. 
    \smallskip
    \item[\textbullet] From DESI data, \cite{Hang2021} reported no significant detection of the stacked ISW signal from voids, and found $A_{\rm ISW}\approx 1.52\pm0.72$ for superclusters, consistent with no evidence for excess signals.
\end{itemize}

Beyond stacking methods using hundreds or thousands of voids, the CMB \emph{Cold Spot}, one of the most extreme anomalies in the CMB, \citep[see e.g.,][]{CruzEtal2004} and the Eridanus supervoid \citep{SzapudiEtAl2014,FinelliEtal2014,KovacsJGB2015} have also been studied extensively. It has been hypothesized that a large void in the LOS could potentially explain the temperature depression at the Cold Spot \citep{RudnickEtal2007}, but in a $\Lambda$CDM model, there is no real chance to explain the total temperature pattern as an ISW signal \citep[see e.g.,][]{Nadathur2015,Naidoo2016,Mackenzie2017}. Nonetheless, \cite{Kovacs2022} provided evidence for a significant supervoid at $z<0.2$ aligned with the Cold Spot, using not only galaxy counts but also weak lensing maps from DES Year-3 data set \citep{Jeffrey2021}, suggesting a correlation.

As additional probes of the CMB imprints of voids and superclusters, a series of analyses have also measured their CMB lensing convergence ($\kappa$) signals \citep[see e.g.,][]{Cai2017,Raghunathan2019, Vielzeuf2019,Hang2021,Kovacs2022_DES,CamachoCiurana2024}, and reported no significant tensions. A state-of-the-art analysis using DESI luminous red galaxies (LRGs) detected the CMB $\kappa$ imprint of voids at the $14\,\sigma$ significance level ($A_{\kappa}\approx 1.016\pm0.054$), again finding great agreement with $\Lambda$CDM \citep{Sartori2024}.

Further extending the redshift range of the ISW measurement, the following results added intriguing threads to the ISW puzzle:
\begin{itemize}
    \item[\textbullet] \cite{Kovacs2021} used an eBOSS QSO sample for ISW stacking measurements at high redshifts, and reported $2.7\,\sigma$ tension with $\Lambda$CDM, and an apparent sign-change at $z>1.5$ ($A_{\rm ISW}\approx 3.6\pm2.1$ at $0.8<z<1.2$, and $A_{\rm ISW}\approx -8.49\pm4.4$ at $1.9<z<2.2$). 
    \smallskip
    \item[\textbullet] More recently, \cite{Hansen:2025atx} reported a sign-change of the ISW signal in the very local Universe ($z < 0.03$), i.e., finding hot imprints for local voids and cold imprints for local over-densities. 
\end{itemize}

These recent indications of (moderately significant) excess signals, and especially the claims for sign-reversal for the ISW effect in some redshift ranges, if confirmed, would demand careful theoretical modeling to exclude foregrounds or selection effects. Resolving these tensions and anomalies, and understanding how ISW signals vary with redshift and scale, requires advances in both observational data and theoretical modelling. On the data side, technological and resource limits impose an upper limit on how much improvement is feasible \citep[see e.g.,][]{Afshordi_2004}. Although upcoming surveys such as \emph{Euclid} are highly promising \citep{euclid_2025}, fully exploiting their datasets will require substantial time and financial investment. 

From the modelling perspective, large-volume simulations and efficient pipelines for generating full-sky ISW maps and mock catalogues are crucial \citep[see e.g.,][]{Watson2014, Beck2018, flagship_2025}. They enable validation of analysis methods, provide realistic covariance estimates, and supply predictions across a broad range of cosmological models. In this work, we use the \emph{Gower Street} (GS, hereafter) suite of simulations, based on the $w$CDM model \citep{jeffrey2024}, to comprehensively study the ISW effect. Our main objective is to create a novel pipeline to reconstruct ISW maps for various cosmologies in GS, enabling \textbf{(i)} precise predictions for the redshift-dependent ISW imprint across a broad range of $w$ from phantom to quintessence domains, \textbf{(ii)} various cross-correlation measurements including 2-point functions, stacking, and simulation-based inference.

The paper is organized as follows. In \autoref{sec:methods}, we describe our methodology and the implementation of the ISW calculations and map-making. In \autoref{sec:results}, we present our main results, which are then further discussed in \autoref{sec:discussion}.

\section{Methodology and implementation}
\label{sec:methods}

%%%%%%%%%%%%%%%%%%%%%%%%%%%%
%%%%%%%%%% Sim %%%%%%%%%%%%%
%%%%%%%%%%%%%%%%%%%%%%%%%%%%
\subsection{Simulation Suite}

The GS simulation suite consists of 791 full-sky dark matter-only simulations run with \pkdgrav{} \citep{Flagship}, based on a flat $w$CDM cosmology spanning $-1.79 \lesssim w \lesssim -0.34$. Each simulation uses a box size of $L = 1250 \, h^{ - 1} \mathrm{Mpc}$, with $N = 1080^3$, starting from $z_0 = 49$. Outputs consist of $100$ light cone files, equally spaced in proper time between $z_0$ and $z = 0$, at a \healpix{}\footnote{\url{https://healpix.sourceforge.io/}} resolution of NSIDE $= 2048$ \citep{healpix}. We note that each GS simulation uses a distinct random seed for the initial condition generation, enabling tests of cosmic variance effects. The simulation data are publicly available \footnote{\url{www.star.ucl.ac.uk/GowerStreetSims/}}. 

\begin{figure}
\centering
\includegraphics[width=90mm]{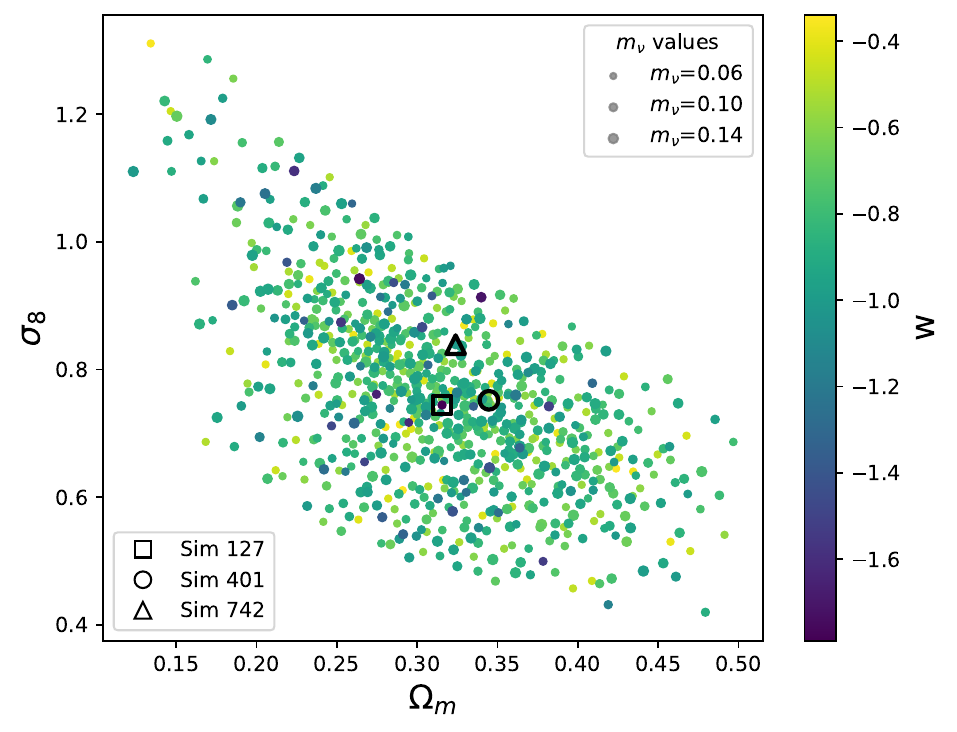}
\caption{Distribution of the 791 GS simulations in an $\Omega_{\mathrm{m}}-\sigma_8$ parameter space, color-coded by $w$, and marker sizes illustrate relative differences in neutrino mass. These parameters are the most relevant for ISW calculations, and the three representative simulations are highlighted with unfilled black markers.}
\label{fig:GS_params}
\end{figure}

The simulations vary seven cosmological parameters: the matter density parameter $\Omega_{\mathrm{m}}$, the amplitude of fluctuations $\sigma_8$, the scalar spectral index $n_s$, the dimensionless Hubble constant $h = H_0 / 100 \, \mathrm{km\, s^{-1}\, Mpc^{-1}}$, the physical baryon density $\Omega_{\mathrm{b}} h^2$, the dark energy equation-of-state (EoS) $w$, and the total neutrino mass $m_{\mathrm{\nu}}$. \autoref{fig:GS_params} illustrates the spread of the GS simulation suite in the $\Omega_{\rm m} - \sigma_{\rm 8}$ parameter space. Each point corresponds to one of the 791 simulations, colour-coded by the dark energy EoS parameter $w$, and with marker size proportional to the total neutrino mass, $m_{\rm \nu}$. 
In what follows, we focus on three representative simulations spanning our parameter space (see \autoref{tab:cosmo_params} for details): 
\begin{itemize}
    \item \textbf{Sim 127}, corresponding to the most extreme phantom dark energy model ($w \simeq -1.79$),
    \item \textbf{Sim 401}, highlighting the strongest quintessence dark energy case in GS suite ($w \simeq -0.34$),
    \item \textbf{Sim 742}, an example for a baseline $\Lambda$CDM model ($w = -1$).
\end{itemize} 
In \autoref{fig:GS_params}, these examples are highlighted with black, unfilled symbols (square, circle, and triangle, respectively), and highlight the broad coverage of the $w$CDM parameter space in the GS simulation suite.

\begin{table*}[t]
\centering
\caption{Cosmological parameters for three representative GS simulations, listing additional parameters besides $\Omega_{\rm m}$, $\sigma_8$, and $w$, which are the key variables in ISW calculations.}
\label{tab:cosmo_params}
\renewcommand{\arraystretch}{1.3}  
\begin{tabular}{c c c c c c c c c}
\hline
Simulation & $\Omega_{\rm m}$ & $\sigma_8$ & $w$ & $h$ & $n_s$ & $m_\nu$ & $\Omega_{\rm b}$ \\
\hline \hline 
{\bf Sim 127} & 0.315 & 0.745 & -1.789 & 0.656 & 0.978 & 0.060 & 0.052 \\
{\bf Sim 401} & 0.345 & 0.752 & -0.339 & 0.700 & 0.962 & 0.072 & 0.046 \\
{\bf Sim 742} & 0.324 & 0.839 & -1.000 & 0.713 & 0.964 & 0.088 & 0.044 \\
\hline
\end{tabular}
\end{table*}

%%%%%%%%%%%%%%%%%%%%%%%%%%%%
%%%%%%%%%% Maps %%%%%%%%%%%%
%%%%%%%%%%%%%%%%%%%%%%%%%%%%
\subsection{Integrated Sachs-Wolfe map construction}
\label{sec:maps}
To generate the ISW temperature maps, we start from the simulated GS light cone catalogues. We use the \texttt{pyGenISW}\footnote{\url{https://github.com/knaidoo29/pyGenISW/tree/master/pyGenISW/isw}} algorithm for the reconstruction of the ISW sky maps from the matter density fluctuations along the line of sight \citep[see][for further detail]{Naidoo2021}. The raw particle counts in each \texttt{HEALPix} pixel are converted into density contrast maps for each light cone slice, which serve as the input for \texttt{pyGenISW}. 

In addition to the tomographic density maps, we constructed a single density contrast map by merging all simulation light cone slices up to $z \simeq 1.0$. Later, we use this projected density map for validating our modelling pipeline using galaxy (particle) auto-power spectra, and also for computing the ISW-density cross-power spectra.

The \texttt{pyGenISW} algorithm performs a spherical Bessel transform based on a ray-tracing calculation of the ISW temperature anisotropies along the line of sight. We highlight that we created an \emph{extended} version of the open-source, $\Lambda$CDM-compatible version of \texttt{pyGenISW}, in order to properly handle $w$CDM models and massive neutrinos in GS simulations. In this work, we focus on the linear ISW effect; non-linear contributions, while possible to include via ray tracing, typically modify the signal by only $\sim\,10\,\%$ \citep[see e.g.][]{Cai2009}. Using these additions, we generated ISW maps up to $z \simeq 1.0$, where 
\begin{itemize}
    \item[\textbullet] the ISW signal is expected to be strongest in most of the models that we are considering.
    \item[\textbullet] possible distortions from simulation box repetition effects are expected to be negligible.
    \item[\textbullet] we expect to have observational data from galaxy surveys to accurately measure these effects in the near future.
\end{itemize}

%%%%%%%%%%%%%%%%%%%%%%%%%%%%
%%%%%%%%%% ISW %%%%%%%%%%%%%
%%%%%%%%%%%%%%%%%%%%%%%%%%%%
\subsection{ISW effect in $w$CDM models}
\label{sec:ISW}
Parallel to the map-based pipeline, we compute theoretical density and ISW power spectra predictions using two independent frameworks: \texttt{TheoryCL}\footnote{\url{https://github.com/knaidoo29/TheoryCL}}, which is part of the original \texttt{pyGenISW} pipeline and we extended to $w$CDM cosmologies as part of this work, and also \texttt{pyCCL}\footnote{\url{https://pypi.org/project/pyccl/}} for validation against a standard, independent code environment \citep{Chisari_2019}. 
For each simulation, the corresponding set of $w$CDM cosmological parameters is fed into both theory pipelines to get the expansion and growth histories.

Our starting point is a simple extension of the $\Lambda$CDM model, referred to as the $w$CDM model. It is described by the same six free parameters as the $\Lambda$CDM model, with one additional parameter accounting for the total mass of massive neutrinos, and with the dark energy EoS parameter allowed to vary ($w_{\rm DE} \equiv w \neq -1$). See \cite{Carbone2016} for further details about the role of $w$ and neutrinos in ISW calculations.

This modifies the Hubble rate through the first Friedmann equation
\begin{equation}
    E^2(a) \equiv \dfrac{H^2(a)}{H_{\rm 0}^2} =  \left[ \Omega_{\rm m} \, a^{-3} + \Omega_{\rm DE} \, a^{-3 \, (1+w)} \right],
\end{equation}
where $H_{\rm 0} = 100\, h \, {\rm km}\, {\rm s}^{-1}\, {\rm Mpc}^{-1}$ is the Hubble constant, $a$ is the scale factor, and $\Omega_{\rm m}$ and $\Omega_{\rm DE}$ are the matter and dark energy density parameters, respectively. The physical CDM density parameter is obtained as
\begin{equation}
    \Omega_{\rm c} = \Omega_{\rm m} - \Omega_{\rm b} - \Omega_{\rm \nu},
    \qquad
     \Omega_{\rm \nu} = \dfrac{\Sigma m_{\rm \nu}}{93.14 \, h^2},
     \label{Eq:m_nu}
\end{equation}
where $\Omega_{b}$ is the baryon density, and $\Omega_{\rm \nu}$ is the neutrino density parameter \cite{Lesgourgues:2006nd,Mauland:2023eax}. 

The cross-correlation between two statistically isotropic Gaussian fields $X$ and $Y$ can be expressed as
\begin{equation}
    C^{XY}_{\rm \ell} \equiv \langle \Theta^X \, \Theta^Y  \rangle = \int^{\infty}_{\rm 0} k^2 \, \mathrm{d}k \, P(k) \, I^X_{\rm \ell}(k) \, I^Y_{\rm \ell}(k),
\end{equation}
where $P(k)$ is the matter power spectrum today \citep{Manzotti2014, Stolzner2018}, and $I^X_{\rm \ell}(k)$ represents the angular transfer (or kernel) function of the field $X$. In this work, we study the auto- and cross-power spectra between the ISW temperature and the matter distribution from the GS simulation. We begin by describing the ISW temperature anisotropy.

The temperature anisotropy of the late-ISW effect, due to the evolution of the gravitational potential, can be expressed as
\begin{equation}
    \Theta^{ISW} \equiv \dfrac{\Delta T^{\rm ISW}}{T_{\rm CMB}} = \dfrac{2}{c^2} \int^{\tau_{\rm 0}}_{\rm \tau_{\rm LS}} \dfrac{\partial \, \Phi({\bf \hat{n}}\, r, \tau)}{\partial \, \tau} \, \mathrm{d} \tau,
    \label{Eq:Theta-ISW}
\end{equation}
where {$\hat{\bf n}$} is the LOS unit vector, $r$ is the comoving radial distance, so {\bf $\hat{\bf n}$}$r$ represent the 3D comoving position vector. $\tau_{\rm LS}$ is conformal time of the last scattering surface, $T_{\rm CMB} = 2.7255 \, \mathrm{K}$ is the CMB temperature today, and $\Phi$ is the gravitational potential, which satisfies the Poisson equation in Fourier space and in Newtonian gauge as
\begin{equation}
    k^2 \, \Phi(k, z) = - \dfrac{3 H^2_{\rm 0} \, \Omega_{\rm m}}{2 \, a(z)} \, \delta_{m}(k, z).  
    \label{Eq:Poisson}    
\end{equation}
Taking all these into account, the ISW transfer function can be written as
\begin{equation}
    I^{ISW}_{\rm \ell}(k) = \dfrac{3\,H^2_{\rm 0}\,\Omega_{\rm m}}{c^2\,k^2} \int \dfrac{\mathrm{d}}{\mathrm{d}z} \left( \dfrac{D(z)}{a(z)} \right) \, j_{\rm \ell} \left[ k_{\rm \chi}(z) \right] \, \mathrm{d}z.
\end{equation}
Similarly, the matter (in our case, particles) transfer function is given by
\begin{equation}
    I^{g}_{\rm \ell}(k) = \int \dfrac{\mathrm{d}N}{\mathrm{d}z} b(z) \, D(z) \, j_{\rm \ell} \left[ k_{\rm \chi}(z) \right] \, \mathrm{d}z,
\end{equation}
where $j_{\rm \ell}\left[ k_{\rm \chi}(z) \right]$ is spherical Bessel function, $\mathrm{d}N/\mathrm{d}z$ is the redshift distribution function of the sources, $b(z)$ is the linear, scale-independent galaxy bias (in our case $b(z)=1$ because we work with dark matter particles), and $D(z)$ is the linear growth factor.
Although the Limber approximation is available in the original implementation of \texttt{TheoryCL}, it is not used in this study. Instead, we evaluate the full expressions above, retaining the spherical Bessel functions, which is particularly important for the low-multipole regime relevant to the ISW effect.

In practical terms, we calculate a simple combination of the expansion rate, $H(z)$, and growth histories, $D(z)$ and $f(z) = \mathrm{d} \ln{D} / \mathrm{d} \ln{a}$, often referred to as the ISW kernel
\begin{equation}
    \Delta T^{\rm ISW} \sim \dot{\Phi} \sim D(z) \, H(z) \, \left[ 1 - f(z) \right].
    \label{eq:kernel}
\end{equation}
Linear matter density perturbations grow as $\delta(\boldsymbol{r},z) =  D(z)\delta(\boldsymbol{r})$, and
$\nabla^2 \Phi(\mathbf{r}, z) = (3/2) H_0^2 \Omega_{\mathrm{m}} \, \delta(\mathbf{r}, z) / a$.
By combining linear growth and the Poisson equation, it follows that $\dot{\Phi}=-H(z)[1-f(z)]\Phi$.

\begin{figure}
\centering
\includegraphics[width=1\columnwidth]{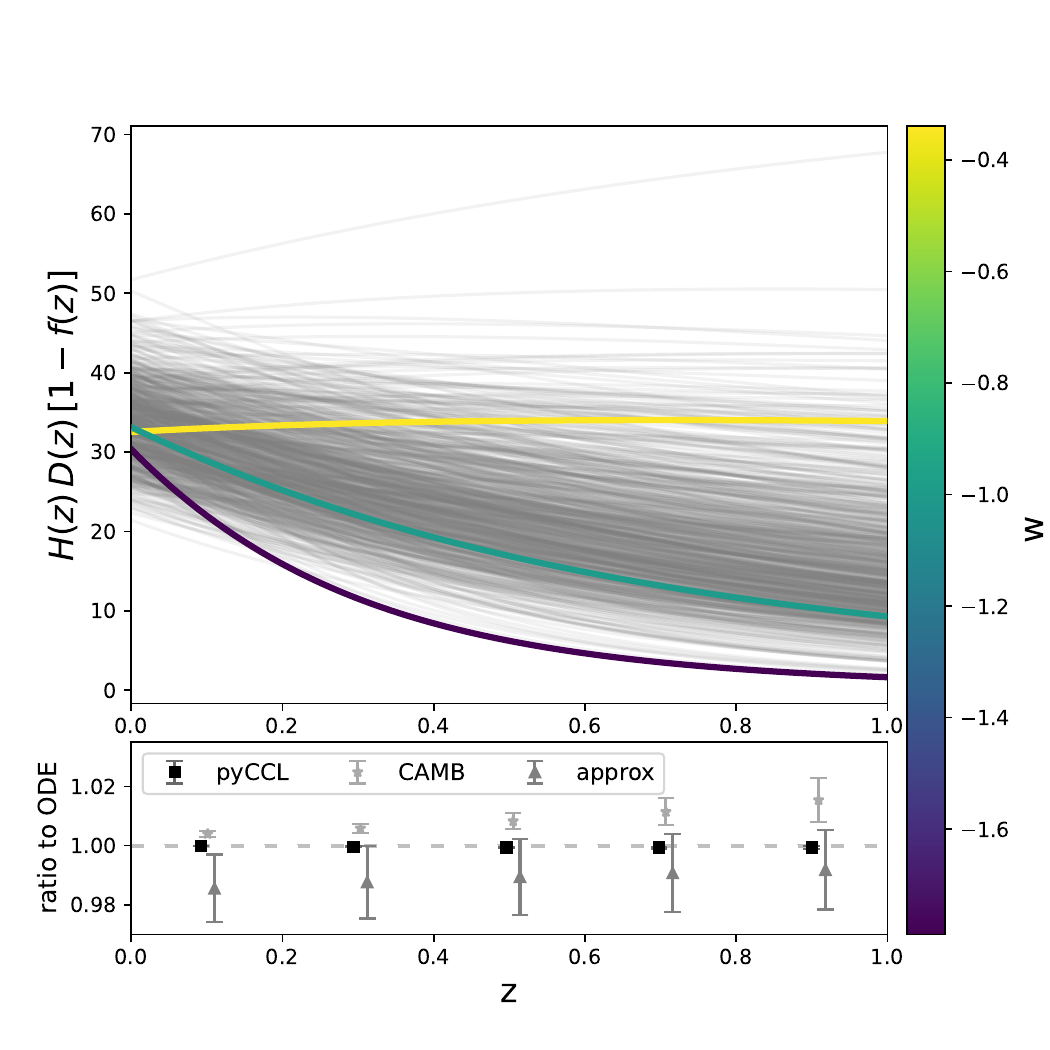}
\caption{{\textbf{Top panel:}} Comparison of the ISW kernel, defined as the product of cosmic expansion and growth, for all the 791 GS simulations (gray). We again highlight simulations 401 (yellow), 742 (green), and 127 (dark blue), as representatives of different $w$ values in the GS parameter space. In the {\textbf{bottom panel}}, we compare our results from the fiducial \texttt{TheoryCL} pipeline and \texttt{pyCCL}, as well as alternative computations using \texttt{CAMB} and approximate $f(z)$ formulae described in \autoref{sec:ISW}. This panel highlights the excellent agreement between our pipeline and  \texttt{pyCCL}.}
\label{fig:kernels}
\end{figure}

%%%%%%%%%%%%%%%%%%%%%%%%%%%
%%%%%%% flowchart %%%%%%%%%
%%%%%%%%%%%%%%%%%%%%%%%%%%%
\begin{figure*}[t]
\centering
\resizebox{\textwidth}{!}{%
\begin{tikzpicture}[
    font=\small,
    box/.style={
        rectangle,
        rounded corners,
        draw=black!60,
        align=center,
        minimum width=3.2cm,
        minimum height=1.1cm
    },
    input/.style={box, fill=black!5},
    model/.style={box, fill=blue!10},
    process/.style={box, fill=blue!10},
    kernel/.style={box, fill=blue!15},
    output/.style={box, fill=blue!20},
    arrow/.style={->, thick, draw=black!70}
]

%%%% inputs %%%%
\node[input] (delta) {
    \textbf{Density contrast maps}\\
    $\delta(\hat{\mathbf n}, z_i)$\\
    light-cone slices \\
    from GS simulations
};

\node[input, below=6mm of delta] (cosmo) {
    \textbf{Cosmological} \\
    \textbf{parameters}\\
    from GS simulations\\
    \\
    $\Omega_{\rm m},\,\Omega_{\rm b},\,h,\,w,$\\
    $n_{\rm s},\,m_{\nu},\,\sigma_8$
};

%%%% TheoryCL %%%%
\node[model, right=8mm of delta] (growth) {
    \textbf{Background \&} \\
    \textbf{Growth Evolution}\\
    extended \texttt{TheoryCL}\\[1mm]
    $\chi(z),\,H(z)$\\
    $D(z)$ (ODE / CAMB)\\
    $f(z)$ (ODE / numerical)
};

%%%% pyGenISW %%%%
\node[process, right=8mm of growth] (alm) {
    \textbf{Harmonic decomposition}\\
    extended \texttt{pyGenISW}\\
    $\delta(\hat{\mathbf n},z_i)\!\rightarrow\!\delta_{\ell m}(z_i)$
};

\node[process, right=8mm of alm] (sbt) {
    \textbf{Spherical Bessel Transform}\\
    $\delta_{\ell m}(z_i)\!\rightarrow\!\delta_{\ell mn}$\\
    exact radial geometry
};

\node[kernel, below=6mm of sbt] (isw) {
    \textbf{ISW kernel}\\
    $\dot{\Phi}(z)$\\
    $\propto D(z)\,H(z)\,[1-f(z)]$
};

\node[output, below=6mm of isw] (almisw) {
    \textbf{ISW prediction}\\
    $a^{\rm ISW}_{\ell m}$
};

%%% ISW map %%%%
\node[
    below=23mm of current bounding box.north,
    anchor=north,
    align=center
] (map) {
    \includegraphics[width=4.3cm]{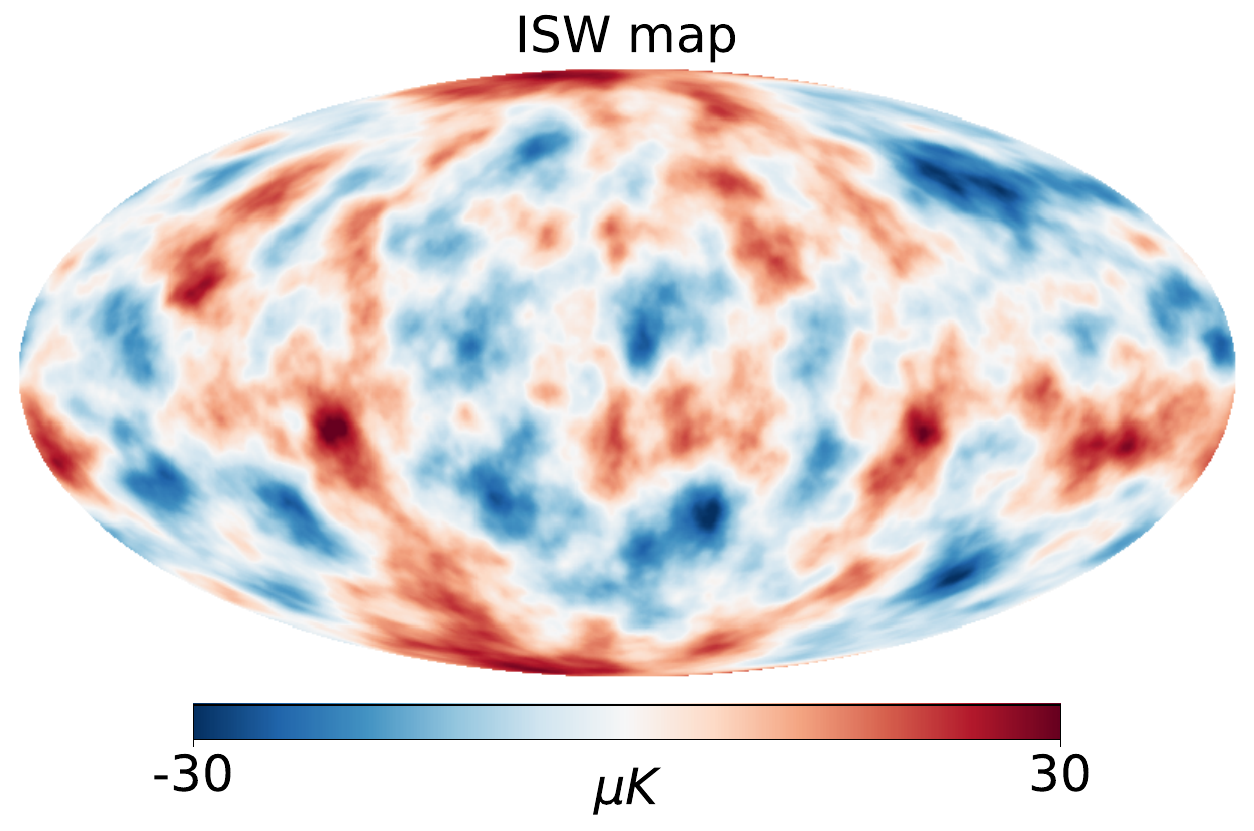}
};

%%%% Arrows %%%%
\draw[arrow] (delta) -- (growth);
\draw[arrow] (cosmo) -- (growth);

\draw[arrow] (growth) -- (alm);
\draw[arrow] (alm) -- (sbt);
\draw[arrow] (sbt) -- (isw);
\draw[arrow] (isw) -- (almisw);

\draw[arrow] (almisw) -- node[above]{\texttt{alm2map}} (map);

%%%% Dashed grouping box %%%%
\node[
    draw=black!50,
    dashed,
    thick,
    rounded corners,
    fit=(growth) (alm) (sbt) (isw) (almisw),
    label={[font=\small]above:{\bf ISW map construction}}
] (modelbox) {};

\end{tikzpicture}
}
\caption{
Simulation-based ISW map construction.
Density contrast maps and cosmological parameters from the Gower Street simulation
are propagated through a unified model (extended \texttt{TheoryCL}) and a
spherical-Bessel ISW pipeline (extended \texttt{pyGenISW}), yielding fully
consistent ISW power spectrum predictions and corresponding ISW temperature maps.
}
\label{fig:isw_pipeline}
\end{figure*}
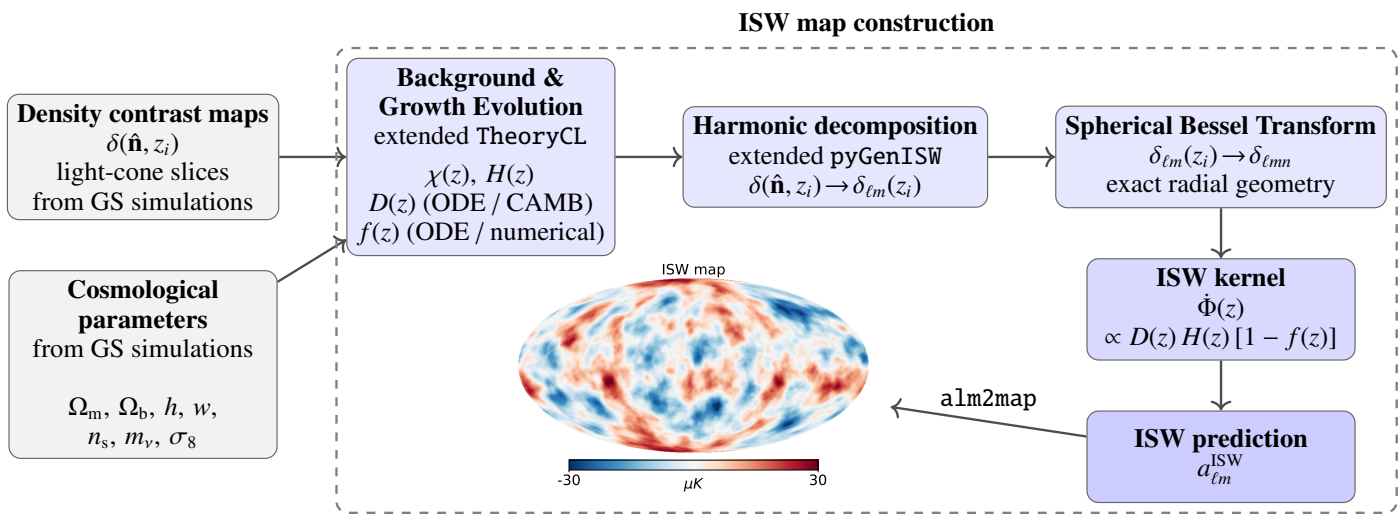

\begin{figure*}
\centering
\includegraphics[width=0.31\textwidth]{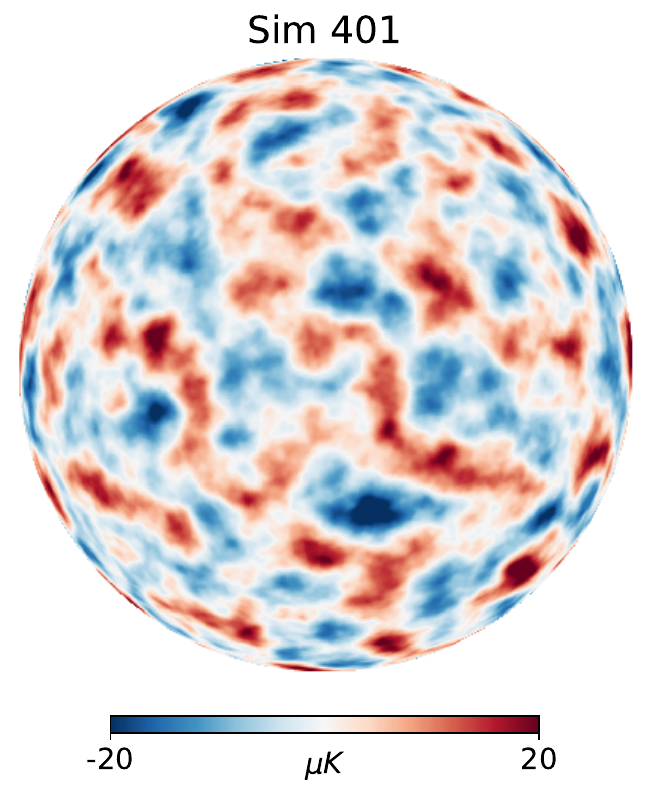}
\includegraphics[width=0.31\textwidth]{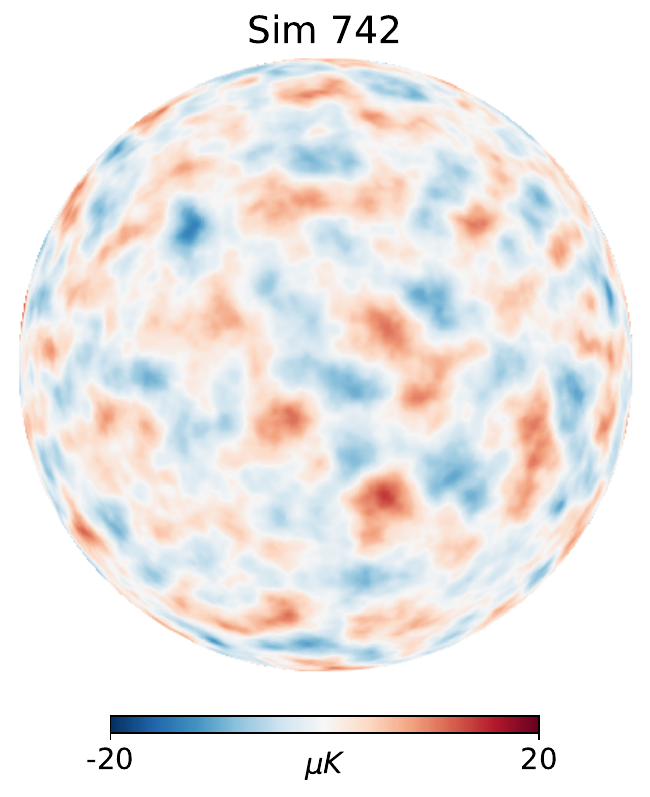}
\includegraphics[width=0.31\textwidth]{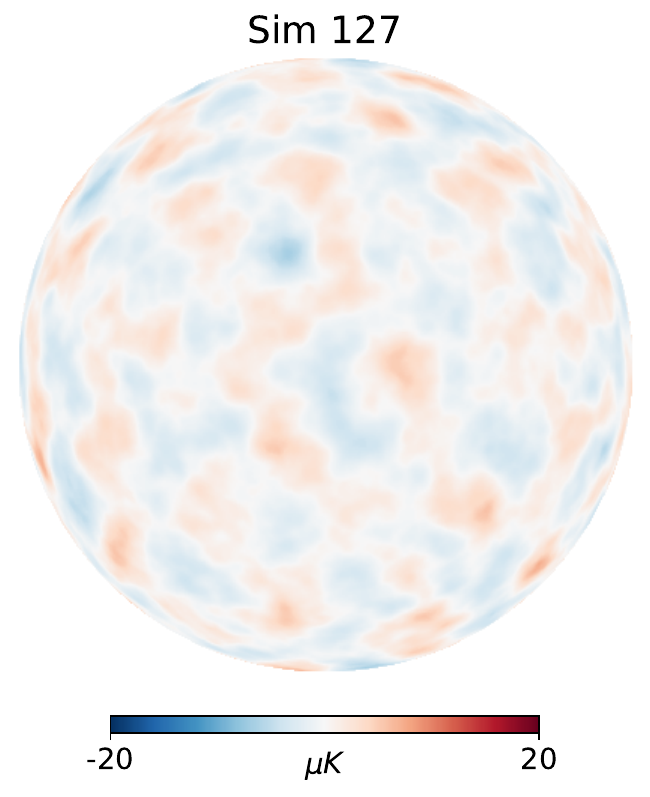}
\caption{Sky-maps of ISW signal, showing half of the sky in an orthographic projection.  Three representatives correspond to simulations 401 ($w=-0.34$), 742 ($w=-1$), and 127 ($w=-1.79$), highlighting different dark energy EOS. The maps are color-coded based on the ISW temperature anisotropies in $\mu K$. The quintessence model shows the most pronounced colors, indicating the strongest ISW signal, followed by the $\Lambda$CDM, while the phantom model shows the weakest signal (see also \autoref{fig:hist_3_ISWmaps} for comparison).}
\label{fig:skyplot_ISWmaps}
\end{figure*}

%%%%%%%%%%%%%%%%%%%%%%%%%%%
%%%%%%%%%%%%%%%%%%%%%%%%%%%

In our pipeline (see \autoref{fig:isw_pipeline}), we implement three complementary methods to compute the linear growth factor and growth rate. Although all three options are available to users, our fiducial choice throughout this work is the ordinary differential equation (ODE)-based solution. 
\begin{itemize}
    \item \textbf{ODE approach (fiducial):} In this method, the linear growth factor is governed by solving the standard differential equation for matter perturbation in a homogeneous and isotropic Universe
\begin{equation}
    \dfrac{\mathrm{d}^2}{\mathrm{d} a^2} D_{\rm ODE} + \dfrac{1}{a} \, \left( 3 + \dfrac{\mathrm{d} \ln{H}}{\mathrm{d} \ln{a}} \right) \, \dfrac{\mathrm{d}}{\mathrm{d} a} D_{\rm ODE} = \dfrac{3}{2 a^2} \, \Omega_{\rm m}(a) \, D_{\rm ODE}(a).
    \label{Eq:ODE}
\end{equation}
This is the baseline approach adopted throughout the paper, and we find excellent agreement with the \texttt{pyCCL} predictions for the GS simulation (which also uses an ODE method for growth calculations). Once $D_{\rm ODE}(z)$ is obtained, the linear growth rate is computed as
\begin{equation}
    f_{\rm ODE}(z) \equiv \dfrac{\mathrm{d} \ln{D_{\rm ODE}}}{\mathrm{d} \ln{a}}.
\end{equation}
In this method, the effect of massive neutrinos enters solely through the background term of $\Omega_{\rm \nu}$ defined in \autoref{Eq:m_nu}, while their scale-dependent clustering is not included in the perturbation equation. We note that in the linear regime where the ISW effect is strongest, and for neutrino mass ranges considered in the GS simulations, the expected additional ISW temperature perturbations by neutrinos are negligible \citep[see e.g.,][]{Cuozzo_2024}.

\, 

    \item \textbf{CAMB-based approach:} As an alternative to solving the growth equation directly, we obtain the linear growth factor from the Boltzmann code \texttt{CAMB}\footnote{\hyperlink{https://camb.readthedocs.io/}{https://camb.readthedocs.io/}}. It computes the full linear matter power spectrum, $P_{\rm lin}(k, z)$, including the scale-dependent suppression induced by massive neutrinos. To extract the growth factor, we follow the standard convention of using the redshift evolution of the root-mean-square (RMS) of linear matter density fluctuations on $8\,h^{-1}{\rm Mpc}$ scales,
    \begin{equation}
        D_{\rm \texttt{CAMB}}(z) \equiv \dfrac{\sigma_{\rm 8}(z)}{\sigma_{\rm 8}(0)}.
    \end{equation}
Here $\sigma_8(z)$ is obtained by integrating the linear matter power spectrum over wavenumber, yielding an effective, scale-averaged amplitude of fluctuations. Even though $P_{\rm lin}(k,z)$ is scale dependent, the resulting $D_{\rm \texttt{CAMB}}(z)$ remains scale independent. In our implementation of this method, the growth rate, $f(z)$, is not provided directly by \texttt{CAMB}; instead, the pipeline computes it numerically, by differentiating the tabulated $D(z)$ values, evaluating $f(z) = \mathrm{d} \ln{D} / \mathrm{d} \ln{a}$ directly from the discrete $D(z)$ grid.

\, 

    \item \textbf{Analytic approximation approach:} For fast evaluations, the pipeline also includes the well-known growth-index approximation, in which the growth rate is parametrized as
\begin{equation}
    f_{\rm approx} \approx \Omega_{\rm m}(z)^{\gamma(w)},
\end{equation}
where $\gamma(w)$ is the growth index \citep{Linder_2005}. In the context of $w$CDM cosmologies, it is approximated by
\begin{equation}
    \gamma(w) = \dfrac{3(1 - w)}{5 - 6w}.
\end{equation}
    This approximation is computationally efficient, but its accuracy degrades for models when $w$ deviates significantly from $-1$, or in models with more complex growth behaviour. Users should therefore verify its suitability for their parameter space. In this approach, the pipeline computes the growth factor using the \texttt{CAMB}-based method. This hybrid approach ensures a stable and accurate $D(z)$ while retaining the speed of the analytic growth-rate approximation.
\end{itemize}

%%%%%%%%%%%%%%%%%%%%%%%%%%%%
%%%%%%%%%% SBT %%%%%%%%%%%%%
%%%%%%%%%%%%%%%%%%%%%%%%%%%%
\subsection{Mapmaking with \texttt{pyGenISW}}

Given the expansion and growth functions, the next step in the pipeline is to construct the underlying 3D density field $\zeta(\chi, \theta, \phi)$ to compute the ISW temperature anisotropies. The density field is expanded in a basis that is adapted to the spherical geometry of the light cone with a finite radial range, where $\chi$ denotes the comoving distance from the observer. 

In this study, we employ the discrete spherical-Bessel transform (SBT) using \texttt{pyGenISW}, which provides an orthonormal set of radial eigenfunctions on a finite space $\chi \in [0, \chi_{\rm max}]$. In this domain, the free eigenfunctions are $j_{\rm \ell}\left[ k_{\rm \ell n} \, \chi \right]$, directly analogous to the $j_{\rm \ell}\left[ k_{\rm \chi}(z) \right]$ terms appearing in the ISW and matter transfer functions. We impose a Neumann boundary condition at the edge of the light cone,
\begin{equation}
    \dfrac{\mathrm{d}}{\mathrm{d}\chi} j_{\rm \ell} (k_{\rm \ell n} \, \chi) \bigg|_{\chi = \chi_{\rm max}} = 0,
\end{equation}
which defines the allowed discrete radial modes 
\begin{equation}
    k_{\rm \ell n} = \dfrac{q_{\rm \ell n}}{\chi_{\rm \max}},
\end{equation}
where $q_{\rm \ell n}$ is the $n$-th zero of $\partial_{\rm x} j_{\rm \ell}(x)$.

In practice, each simulated \healpix{} density-contrast slice is first converted into spherical-harmonic coefficients, $a_{\rm \ell m}$, via a \healpix{} transform (using \texttt{alm2map}). These coefficients are then projected into the discrete SBT basis to obtain the radial coefficients, $\delta_{\rm \ell m n}$ on the finite radial intervals. For each slice, we introduced a LOS-weighted effective radial position, applied the corresponding linear growth correction to that slice, and then evaluated the projection factor, $S_{\rm \ell n}$. Radial normalization, $N_{\rm \ell n}$ follow the analytic expression for the chosen Neumann boundary condition.

We thus calculate the ISW spherical-harmonic coefficients by integrating the density field against the spherical Bessel functions for each ($\ell, n$) mode, summing over all discrete radial modes to obtain the required $a_{\rm \ell m}$ for ISW. These $a_{\rm \ell m}$ are then converted into a temperature anisotropy map with an inverse spherical harmonic transform. 

To ensure consistency with the map-making pipeline, all theoretical calculations use the galaxy (in our case, particle) redshift distribution, $N(z)$, measured directly from the simulation lightcones. From these maps and kernels, we compute the ISW auto-power spectrum, the density auto-power spectrum, and the ISW-density cross-power spectrum.

Computationally, this pipeline follows the method introduced by \cite{Naidoo2021}; extending it with additional cosmological parameters as input arguments, a user-selected growth function, and a factor prescription. We also modified the radial window function normalization, created alternative treatments of slice-effective radii, improved numerical integration routines (\texttt{Simpson} instead of \texttt{trapezoidal}), and enhanced the spherical harmonic pipeline that incorporates pixel-window deconvolution, which is handled within the \texttt{pyGenISW}-based code.

%%%%%%%%%%%%%%%%%%%%%%%%%%%%
%%%%%%%% Results %%%%%%%%%%%
%%%%%%%%%%%%%%%%%%%%%%%%%%%%
\section{Results}
\label{sec:results}

\subsection{Validation of the theoretical framework}

Before computing the $w$CDM ISW maps, we first validate the theoretical framework by examining the redshift evolution of the ISW kernel in the linear regime. The growth factor and growth rate, obtained from the extended \texttt{TheoryCL} pipeline, feed directly into the ISW kernel, defined in \autoref{eq:kernel} \citep[see e.g.,][]{Cai2009}. This allows us to verify that our pipeline correctly captures the expected dependency on cosmological parameters. 

\begin{figure*}
\centering
\includegraphics[width=180mm]{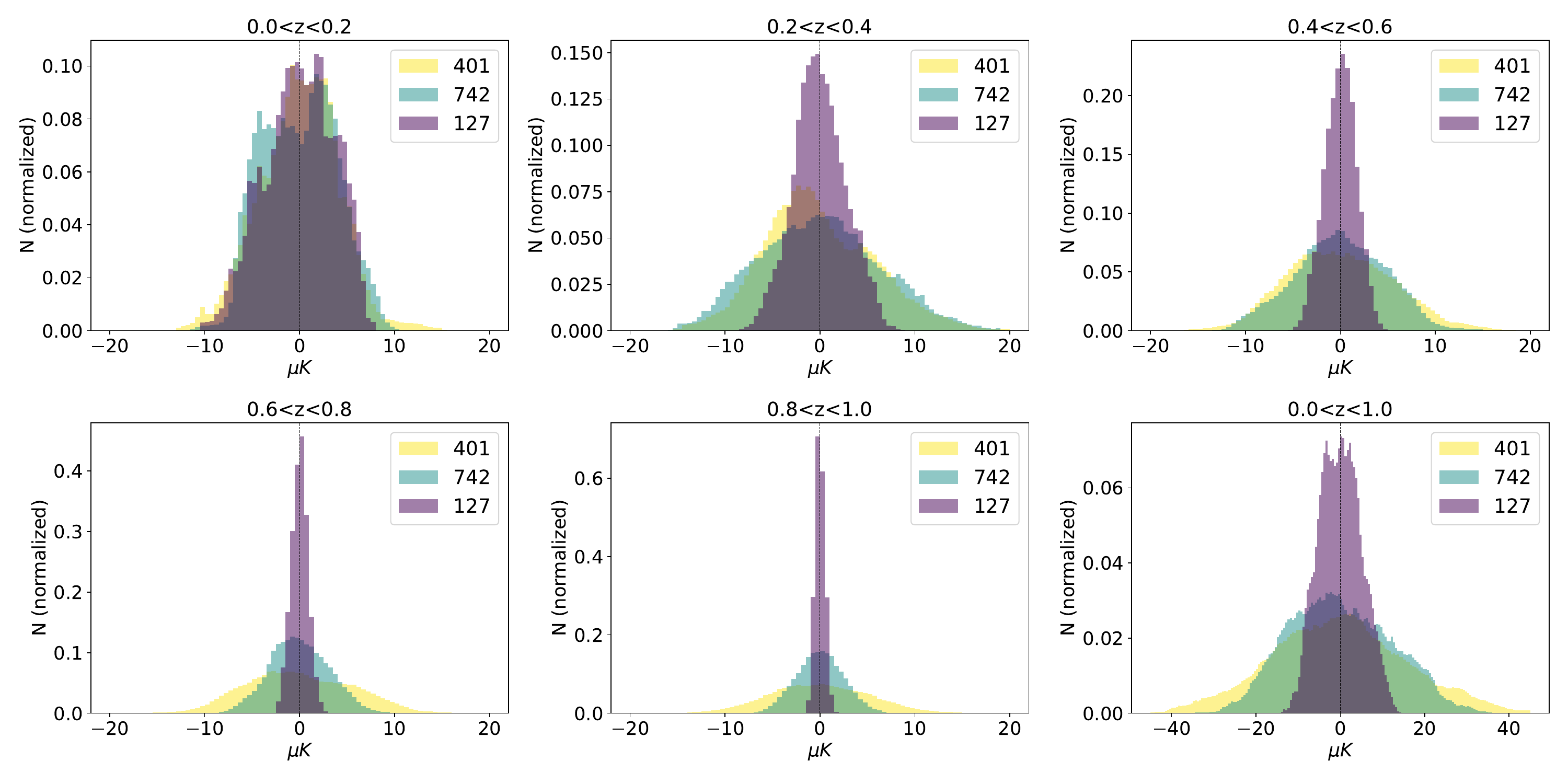}
\caption{Histograms of ISW temperature maps for the three characteristic examples: Sim 127 ($w=-1.79$), Sim 742 ($w=-1$), and Sim 401 ($w=-0.34$). Different panels show redshift-binned ISW temperature statistics, using \healpix{} pixel values. Clear trends are visible with varying $w$ parameters, and also with redshift.}
\label{fig:hist_3_ISWmaps}
\end{figure*}

The top panel of \autoref{fig:kernels} shows the redshift evolution of the ISW kernel in linear theory, where curves cover the interval $0 \leq z \leq 1$, consistent with the redshift range of the ISW maps created in this work (limited by box repetition effects). We display results from all of our ODE-based, extended \texttt{TheoryCL} computations in gray, spanning the full range of cosmologies sampled by the GS simulations, color-coded according to their $w$ values. 

The bottom panel in \autoref{fig:kernels} shows the ratio of several alternative growth-rate prescriptions to the ODE-based calculation. We compare results from \texttt{pyCCL} (black), the \texttt{CAMB}-based growth factor (light gray), and the analytic approximation method (gray). Their ratios are evaluated at five representative redshifts, and each data point represents the mean over 791 simulations, with error bars showing the corresponding standard deviation.

We identified several trends. First, the ODE approach of \texttt{TheoryCL} kernel agrees extremely well with \texttt{pyCCL} across all redshifts, as expected since both packages solve the same underlying differential equation. In contrast, the \texttt{CAMB}-based values show systematic deviations of up to $~ 2\%$ at high redshift, with a growing offset from the central value of unity, and the error bars also increase with $z$. The analytic approximation shows somewhat larger scatter at all redshifts, though its $1\,\sigma$ error bars overlap with unity for $z \gtrsim 0.3$. 

Overall, \autoref{fig:kernels} highlights the robustness and reliability of our \texttt{TheoryCL} ODE approach, yielding ISW kernel predictions that are fully consistent with those from the widely used \texttt{pyCCL} framework for all $w$CDM cosmologies in our sample.

\begin{figure*}
\centering
\includegraphics[width=90mm]{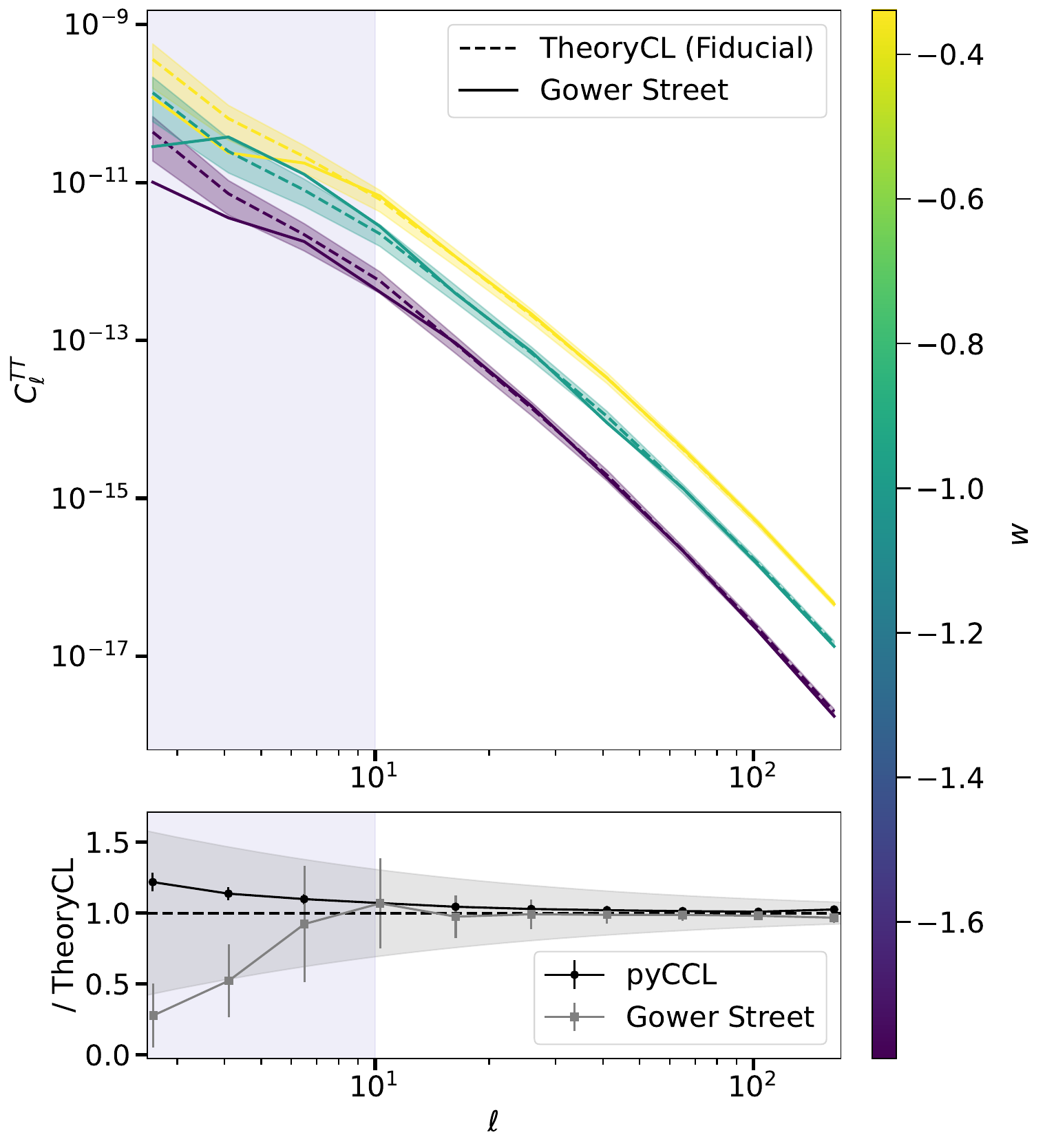}
\includegraphics[width=90mm]{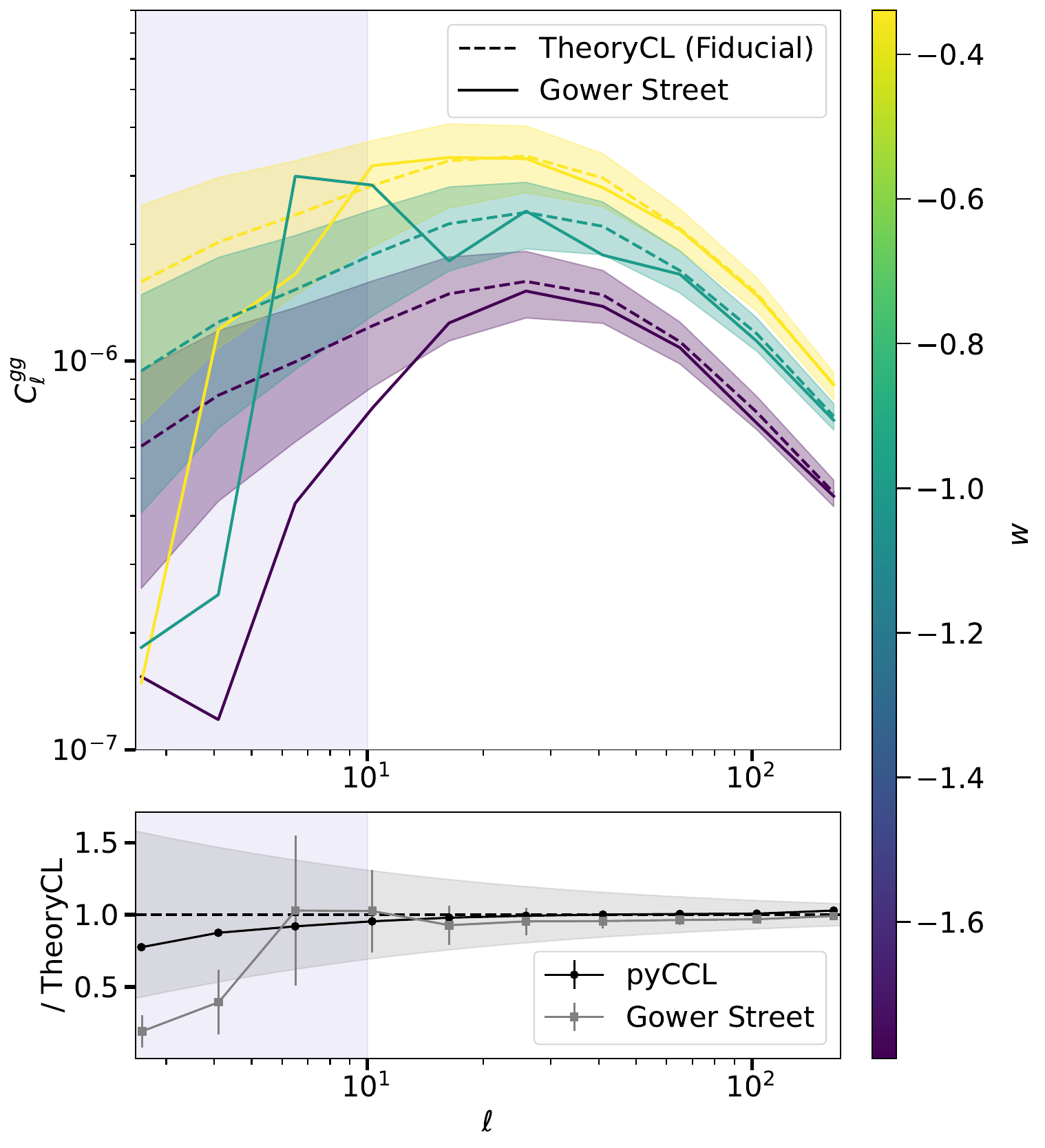}
\caption{Theoretical vs. map-based power spectra for ISW auto-correlations (left) and galaxy auto-correlations (right). The top panels again display the three characteristic models with different $w$ parameters, comparing the measured value from \texttt{pyGenISW} and \texttt{anafast} (solid) with the theoretical expectations from \texttt{TheoryCL} (dashed). We mark the limitations due to cosmic variance (shaded about dashed lines), and also the $\ell < 10$ range where we expect to have imperfect density and ISW maps due to the GS box size limitations. In the bottom panels, we provide statistical results for the agreement of the fiducial \texttt{TheoryCL} and alternative \texttt{pyCCL} theoretical pipelines, and also a comparison between theory and map-based results for the ensemble of the GS mocks. We report good agreement, with a slight deficit of power at small scales ($\ell \geq 100$) where the ISW signal is weakest, and where linear theory assumptions might break down first.}
\label{fig:CISW_summary}
\end{figure*}

\subsection{ISW map statistics}

For 777 of the 791 GS mocks, we managed to create an ISW map using our extended \texttt{pyGenISW} pipeline. In the remaining 14 simulations, we identified numerical instabilities or convergence issues that prevented reliable ISW map generation.
In \autoref{fig:skyplot_ISWmaps}, we show three representative ISW maps generated with our pipeline. As expected, the overall ISW fluctuation power is strongest for \textbf{Sim 401} with $w=-0.34$ (determined by a consistently strong kernel values up to high redshifts, see again \autoref{fig:kernels}). In line with our expectations, \textbf{Sim 742} ranks second in the ISW signal strength, and then \textbf{Sim 127} features the weakest overall ISW pattern, due to a later onset of the dark energy component with $w=-1.79$. We note that we removed the $\ell<10$ modes from the ISW maps, for better visualization of the largest scales that we can certainly reconstruct with the GS simulations, given their box size.

In \autoref{fig:hist_3_ISWmaps}, we also illustrate the ISW signal contributions from different redshift ranges in these three simulations with different $w$ values. At low redshift ($z < 0.2$), all models have nearly identical histograms despite their remarkably different values of $w$. But we should consider that $w$ is not the only relevant cosmological parameter here (see \autoref{tab:cosmo_params} for a complete list), especially when it comes to the low-redshift regime. In this study, we calculate the ISW map fully in the linear regime, and the growth rate of linear gravitational potentials is already low at the present epoch. 
This is controlled primarily by $\Omega_{\rm m}$, which in our three simulations varies only up to $\sim 8\%$ level between \textbf{Sim 127} and \textbf{Sim 401}. On the other hand, \textbf{Sim 742} has the highest $\sigma_{\rm 8}$ around $10\%$ higher than \textbf{Sim 127}, which can compensate for the differences caused by $w$. As a result, the models become distinguishable only at intermediate redshifts where the ISW kernel reaches its maximum sensitivity to the growth history. We made the following observations:

\begin{itemize}
    \item[\textbullet] From $z \approx 0.3$ onward, differences become more visible, when the phantom model (\textbf{Sim 127}) generally produces narrower and sharper distributions, indicating smaller ISW signals, while the quintessence model (\textbf{Sim 401}) shows broader and shallower distributions, reflecting larger temperature fluctuations. 
    \smallskip
    \item[\textbullet] The strongest separation occurs around $0.6 < z < 1.0$. Because the horizontal axis shows temperature fluctuations in $\mu K$ around zero, a sharper histogram corresponds to a weaker overall signal. In this intermediate interval, the quintessence model (\textbf{Sim 401}) shows the broadest distribution, showing the strongest ISW fluctuations among the three representatives in these redshift bins, consistent with the fact that in a quintessence-like cosmology, matter-$\Lambda$ equality occurs at a higher redshift (earlier time) than in phantom models for the same present-day densities. 
    \smallskip
    \item[\textbullet] In the final panel for the entire redshift interval that we considered in GS, $0.0 < z < 1.0$, the ordering of peaks with the highest for the phantom model reflects the cumulative effects induced by different dark energy EoS.
\end{itemize}

\begin{figure}
\centering
\includegraphics[width=90mm]{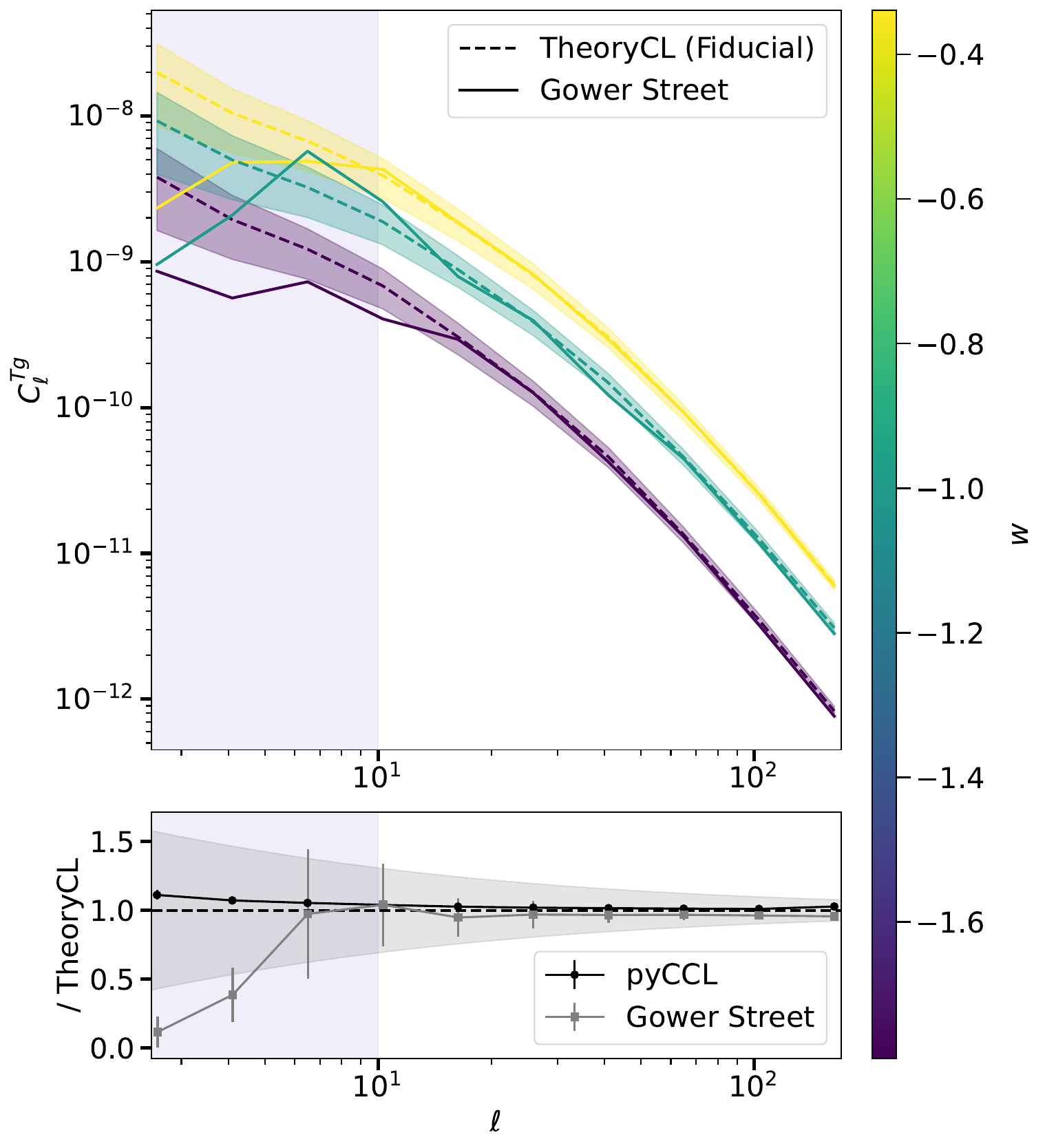}
\caption{Similar to \autoref{fig:CISW_summary}, but for the galaxy-ISW cross-correlation. Different simulations with varying $w$ parameters show different power spectrum amplitudes, and the overall agreement is great between theory and map-based results, except for the lowest and highest $\ell$-ranges.}
\label{fig:CgT_summary}
\end{figure}

Relatedly, most current and upcoming surveys (e.g., \emph{Euclid}, Rubin-LSST, DESI) provide their highest signal-to-noise measurements of galaxies in this redshift range. Our simulated ISW maps will provide a new environment for covariance estimation and higher-order statistical modelling tools to fully extract the joint LSS-CMB information from these surveys.

\subsection{Simulation-based vs. theoretical power spectra}

In \autoref{fig:CISW_summary}, we show the ISW auto-power spectrum, $C^{\rm ISW-ISW}_{\rm \ell}$ in the left panel, over the multipole range $2 \leq \ell \leq 200$, again for the three representative simulations. 
For each model, we compare the theoretical predictions from \texttt{TheoryCL} with the power spectrum measurements using the \texttt{anafast}\footnote{This program performs harmonic analysis of \healpix{} maps up to a user-specified maximum spherical harmonic order $\ell_{\mathrm{max}}$.} method, estimated from our \texttt{pyGenISW}-based ISW maps. Both sets of curves are logarithmically binned to suppress small-scale fluctuations and to emphasize the large-scale behaviors. The color-coded shaded regions indicate the $1\, \sigma$ cosmic variance band derived from the \texttt{TheoryCL} predictions as \begin{equation}
\sigma_{C_\ell} = \sqrt{\frac{2}{(2\ell + 1) f_{\rm sky}}} \, C_\ell
\end{equation}
where $f_{\rm sky}$ is the observed sky fraction (in our case 1) and $C_\ell$ is the angular power spectrum.

Creating a summary statistic for the performance of our ISW map-making pipeline using all the GS simulations, the bottom panel shows the ratio between map-based results from \texttt{pyGenISW} (grey points) and from \texttt{pyCCL} (black points) and the \texttt{TheoryCL} predictions. We cannot expect a perfect match simply due to \emph{cosmic variance} limitations, but we ensured an excellent statistical match for the 777 correctly processed mocks. For reference, the $1\, \sigma$ cosmic variance based on the \texttt{TheoryCL} spectrum is also shown, displaying the expected imperfections for a single map. The points and error bars in the bottom panel represent the mean and standard deviation of the ratios over all 777 simulations, for each of the 10 logarithmic bins.

In both panels, $\ell < 10$ are visually shaded to highlight the limitations imposed by the finite simulation volume. To create a full sky light cone from repeated simulation boxes, modes larger than the box size are not independent, leading to artificial large-scale power and unreliable estimates in these multipoles. Furthermore, we also note that in typical ISW measurements, these modes are often removed from the analysis due to possible biases due to masking, and considering that these scales are much larger than the typical angular size of voids and superclusters \citep[see e.g.,][]{Kovacs2020}.

The right column of \autoref{fig:CISW_summary} shows the galaxy auto-power spectrum, $C^{gg}_{\rm \ell}$, which does not contain any ISW contribution. For the \texttt{pyGenISW}-based measurements, we obtain it directly from the merged density-contrast maps, constructed by stacking all lightcone slices up to $z \simeq 1.0$, and compute the harmonic-space spectrum using \texttt{anafast}. The corresponding theoretical predictions from \texttt{TheoryCL} and \texttt{pyCCL} use the simulated $N(z)$ as explained above. Apart from the physical quantity being plotted, the treatment is identical to the left column. We again find great agreement between theoretical and map-based power spectra, except at $\ell < 10$ where the GS simulations become insufficient to model the matter density fluctuations due to box-size effects accurately.

In \autoref{fig:CgT_summary}, we also show our results for the ISW-galaxy cross-power spectrum statistics. We again report good agreement between the two theoretical pipelines. Theoretical and map-based power spectra are in close match when looking at individual simulations and cosmic variance limits, or when considering the overall ensemble-average statistics for the 777 GS mocks.

Overall, we report great agreement between theoretical expectations and the properties of the ISW temperature maps that we generated for the GS simulations, in an extended $w$CDM parameter space. The largest scales at $\ell < 10$, and the smallest scales at $\ell \geq 100$ are where we find some power deficit in the maps, also in the ensemble averages, given their standard deviations. However, these features in $C^{\rm gg}_{\rm \ell}$, $C^{\rm ISW-ISW}_{\rm \ell}$, and $C^{\rm g-ISW}_{\rm \ell}$ originated from box-size and resolution effects that are expected consequences of our input simulations and our methodology, and they remain at the $\sim1-2\,\sigma$ level.

\subsection{ISW signals in density extrema}

We also tested the role of the input cosmology by splitting the $\delta_{\rm m}$ projected matter density field into percentiles, i.e., identifying troughs and peaks where the ISW is expected to be maximal \citep[see e.g.,][]{Gruen2016}. To reduce small-scale fluctuations in the map, we applied a $\sigma=5^{\circ}$ Gaussian smoothing to the density field, while from the ISW map we again removed $\ell < 10$ modes.

\begin{figure}
\centering
\includegraphics[width=90mm]{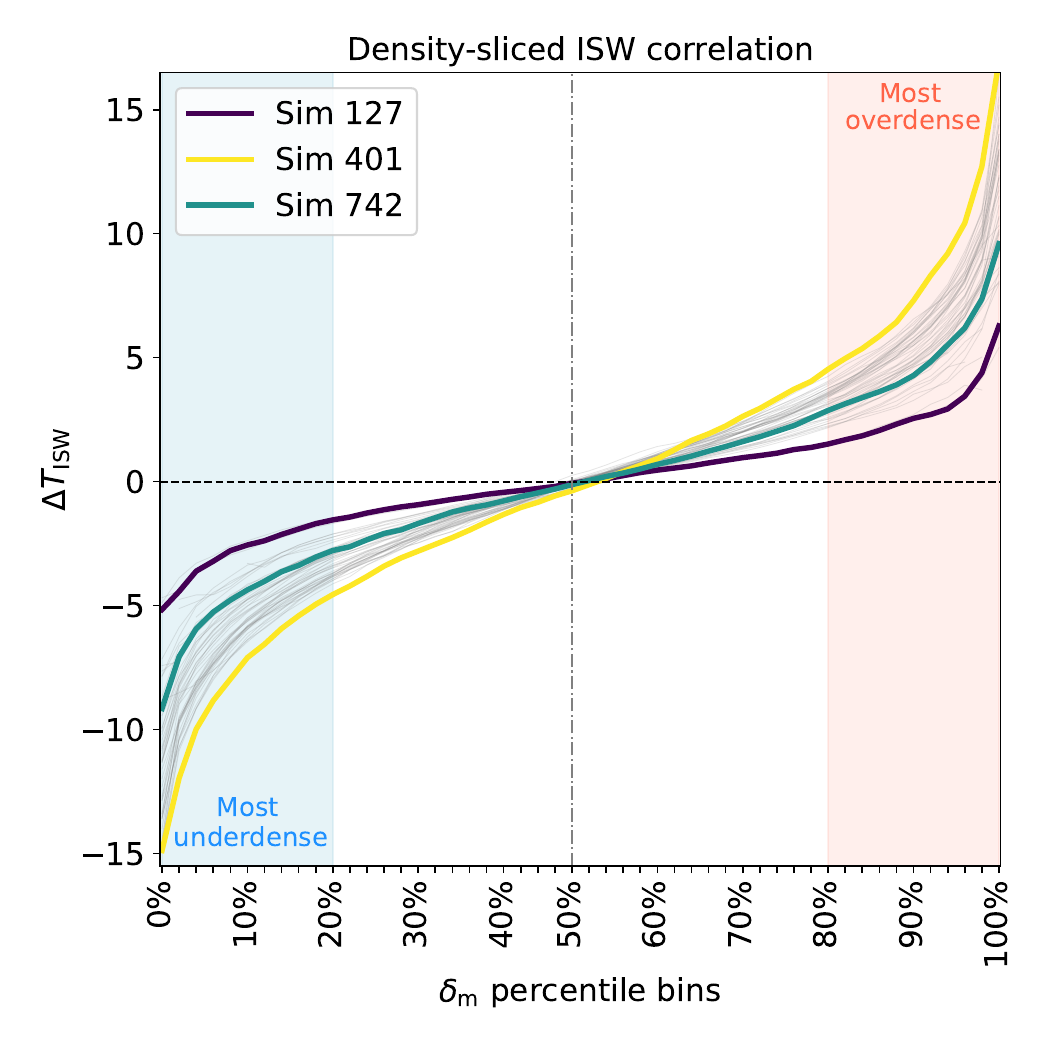}
\caption{Mean $\Delta T_{\rm ISW}$ temperatures in percentiles of the $\delta_{\rm m}$ projected matter density field. The impact of varying the input cosmological parameters is shown here in the most extreme parts of the density field, where the ISW signals differ most (in blue and red shaded areas).}
\label{fig:peak_trough}
\end{figure}

For a random subset of 50 $w$CDM simulations, we measured the mean $\Delta T_{\rm ISW}$ temperatures in each density-split bin defined in the \healpix{} maps. Providing an example for map-level applications of our GS ISW products, these results are summarized in \autoref{fig:peak_trough}. As expected, we found that the magnitude of the ISW signal is enhanced in density extrema, and it shows a clear dependence on the underlying dark energy model. In particular, the quintessence model (\textbf{Sim 401}), shows the largest temperature shifts in both the most overdense and underdense regions, reaching to $\Delta T_{\rm ISW} \sim \pm 15 \, \mu K$, while the phantom model (\textbf{Sim 127}), displays a weaker signal $\Delta T_{\rm ISW} \sim \pm 5 \, \mu K$. This is also consistent with the results shown in \autoref{fig:hist_3_ISWmaps}. When we apply observational data from galaxy surveys, such measurements could provide complementary constraints to traditional two-point statistics, particularly by exploiting the enhanced signal in voids and clusters. 

%%%%%%%%%%%%%%%%%%%%%%%%%%%%
%%%%%%%% A_{ISW} %%%%%%%%%%%
%%%%%%%%%%%%%%%%%%%%%%%%%%%%
\subsection{ISW power ratios in different models}
We further quantified the relative strength of the ISW signal in different models.
As introduced in \autoref{sec:intro}, the $A_{\rm ISW}$ parameter is usually defined as the ratio of the ISW signal measured from data, and a prediction from a fiducial $\Lambda$CDM model (\autoref{eq:A_ISW}). Along similar lines, here we compute for all GS models the following ISW power ratio
\begin{equation}
A_{\rm ISW}^{Tg}
\;=\;
\left\langle
\frac{\bar{C}^{\, \rm Tg,\, \mathrm{sim}}_{\rm \ell}}
     {\bar{C}^{\, \rm Tg, \,\mathrm{fid}}_{\rm \ell}}
\right\rangle_{[\ell_{\min}, \, \ell_{\max}]},
\end{equation}
with cross-power spectrum $C^{\rm Tg}_{\ell}$, using \textbf{Sim 742} as our reference, fiducial $\Lambda$CDM model. Given the limitations of our GS methodology, we calculate the above power ratio in the multipole range $10 \leq \ell \leq 200$.

Therefore, we do not fit a field-level ISW amplitude or power spectrum template, but we instead compress the ISW information into a single scalar quantity, the power ratio $A^{\rm Tg}_{\rm ISW}$, by comparing the overall normalization of the angular power spectra. \autoref{fig:A_ISW} shows the resulting ISW amplitudes, $A^{\rm Tg}_{\rm ISW}$ as a function of the matter density parameter, $\Omega_{\rm m}$, for all 777 valid GS simulations. The points are colour-coded by the dark energy EOS parameter, $w$, and three representative simulations are highlighted with different markers. We made the following observations:

\begin{itemize}
    \item The horizontal dashed line indicates the fiducial $\Lambda$CDM expectation at $A^{\rm Tg}_{\rm ISW} = 1$, among a set of models spanning more than an order of magnitude, reflecting the sensitivity of the ISW effect to late-time cosmological evolution across our $w$CDM parameter space. The inset histogram shows the distribution of $A^{\rm Tg}_{\rm ISW}$. 
    \smallskip
    \item The wide extent of the distributions with about $0.1 \lesssim A^{\rm Tg}_{\rm ISW} \lesssim 10$ reflects the diversity of cosmological models in the simulation suite. In particular, the broad ranges in parameters such as $\Omega_{\rm m}$ and $w$ lead to large variations in the evolution of the late-time gravitational potential.    
    \smallskip
    \item We note that $w$CDM models with $w>-1$ and low values of $\Omega_{\rm m}$ produce ISW amplitudes that are consistent with the observed $A_{\rm ISW}\approx5$ from some low-$z$ voids. However, no model analyzed in this work can explain the apparent sign-change at $z<0.1$ and $z>1.5$ in void stacking analyses.
\end{itemize}

%%%%%%%%%%%%%%%%%%%%%%%%%%%%
%%%%%%% Discussion %%%%%%%%%
%%%%%%%%%%%%%%%%%%%%%%%%%%%%

\section{Discussion and Conclusions}
\label{sec:discussion}
In this work, we developed and validated a novel framework for generating ISW temperature maps for a set of 777 $w$CDM cosmological models in the Gower Street simulation suite \citep{jeffrey2024}. In particular, we modified and extended the $\Lambda$CDM-compatible \texttt{pyGenISW} temperature map reconstruction methodology \citep{Naidoo2021}, allowing accurate and robust modelling of the ISW signal in a wide range of $w$CDM models. Our main results are the following:

\begin{figure}
\centering
\includegraphics[width=90mm]{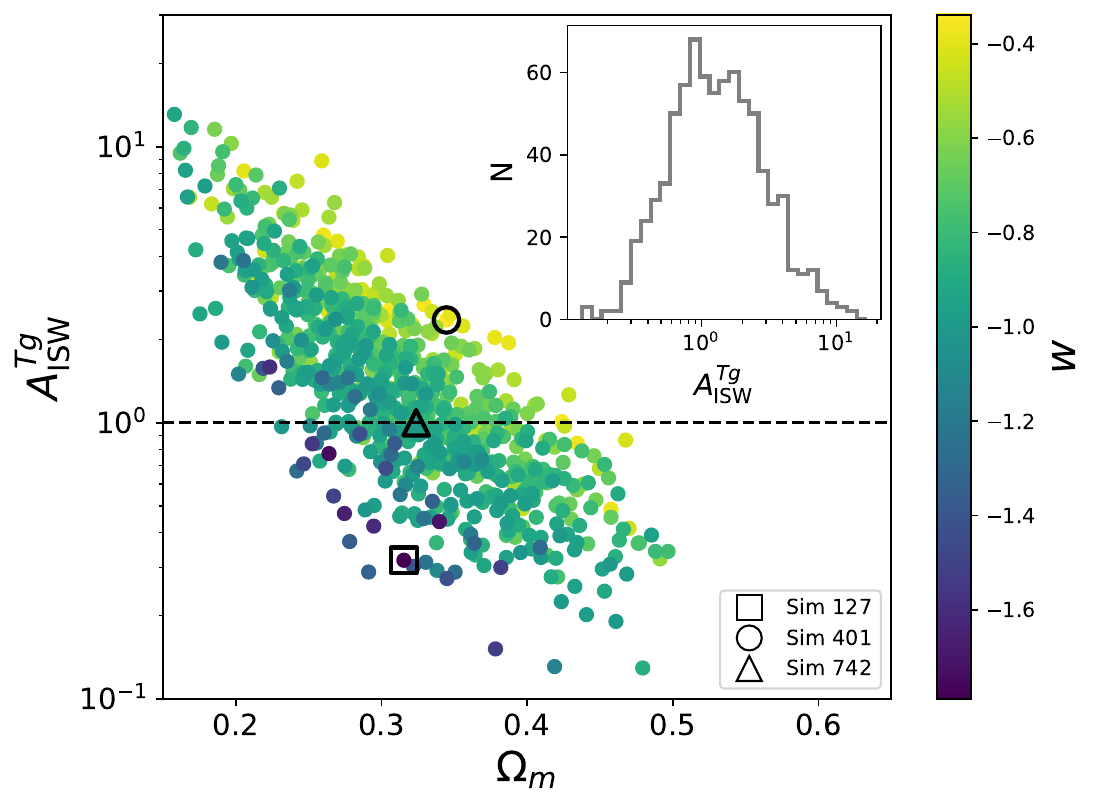}
\caption{
ISW amplitude $A^{\rm Tg}_{\rm ISW}$, computed from power spectra ratios for $777$ simulations spanning a wide range of $w$CDM cosmological parameters. The main scatter plot shows $A_{\rm ISW}^{\rm Tg}$ as a function of $\Omega_{\rm m}$, with points colour-coded by $w$. Three representative simulations (Sim 127, Sim 401, Sim 742) are highlighted with unfilled black markers. The horizontal dashed line indicates the fiducial $\Lambda$CDM expectation $A^{\rm Tg}_{\rm ISW}=1$. The inset histogram shows the distribution of the $A^{\rm Tg}_{\rm ISW}$, also showing a broad variation in ISW strength across cosmological models.}
\label{fig:A_ISW}
\end{figure}

\begin{itemize}
    \item[\textbullet] In our validation tests, we compared our theoretical predictions for $C^{\rm gg}_{\rm \ell}$, $C^{\rm ISW-ISW}_{\rm \ell}$, and $C^{\rm g-ISW}_{\rm \ell}$ auto- and cross-power spectra with multiple alternatives, finding that the ODE approach to modelling the growth factor does provide robust results, consistent with those obtained from the widely adopted \texttt{pyCCL} across the entire $0 \leq z \leq 1$ redshift interval most relevant to ISW. \bigskip
    \item[\textbullet] Using this framework, we analyzed a suite of simulations spanning a wide range of constant $w$ models, including some extreme phantom ($w=-1.79$) and quintessence ($w=-0.34$) cases. The ISW auto-power spectra extracted from our \texttt{pyGenISW} maps show excellent agreement with theoretical predictions, when compared in logarithmic multipole bins over $2 \leq \ell \leq 200$. The residuals lie well within the expected cosmic-variance envelope for all simulations. At the largest angular scales ($\ell < 10$), the discrepancies follow the expected behavior due to the finite volume of the GS simulations, and the associated lack of independent large-scale modes.     
    \bigskip
    \item[\textbullet] We further verified that the $C^{\rm gg}_{\rm \ell}$ galaxy auto-power spectra derived from the simulated density maps are in agreement with \texttt{TheoryCL} (and \texttt{pyCCL}) predictions when adopting the corresponding $N(z)$ distributions of the GS simulations.
    \bigskip
    \item[\textbullet] The ISW temperature map histograms evaluated in redshift bins show that all three highlighted cosmologies with radically different $w$ parameters ($w=-1.79$, $w=-1$, $w=-0.34$) remain nearly indistinguishable at very low redshift. This finding indicates that the similar present-day decay rate of gravitational potentials is mostly set by their $\Omega_{\rm m}$ values. More noticeable differences only emerge at $z\geq0.2$, where the ISW signal is more sensitive to the growth history. 
    \bigskip
    \item[\textbullet] However, integration over the full redshift interval that we considered in this GS analysis, up to $z\simeq 1$, distinguishes the three models. We measure the following standard deviations in the ISW maps $\sigma_{\Delta T_{\rm ISW}}\approx 4.99 \, \mu K$ for \textbf{Sim 127}, $\sigma_{\Delta T_{\rm ISW}}\approx 12.12 \, \mu K$ for \textbf{Sim 742}, and $\sigma_{\Delta T_{\rm ISW}}\approx 16.47 \, \mu K$ for \textbf{Sim 401} (see \autoref{fig:hist_3_ISWmaps} for detail). This increased variance in the map as a result of lower $w$ values confirms that the ISW signal has contributions across a wide range of cosmic epochs rather than solely from the very local Universe.
    \bigskip
    \item[\textbullet]
    Testing the extrema, we also split the projected matter density field into percentiles and thus identified extreme troughs and peaks with presumably strong local ISW signals in our maps. As expected, we found that the quintessence model (\textbf{Sim 401}) shows the largest temperature shifts in both the most overdense and underdense regions, while the phantom model (\textbf{Sim 127}) displays a weaker signal than the reference $\Lambda$CDM model (\textbf{Sim 742}).
    \bigskip
    \item[\textbullet]
    Computing power ratios from $C_\ell^{\rm Tg}$, we estimated $A^{\rm Tg}_{\rm ISW}$ for all 777 GS simulations, all compared to a fiducial $\Lambda$CDM model. The resulting scatter, shown in \autoref{fig:A_ISW}, reflects the combined influence of multiple cosmological parameters. While there is no simple monotonic trend with $w$, simulations with comparable $\Omega_{\rm m}$ values suggest that quintessence-like models ($w > -1$) tend to exhibit slightly higher ISW amplitudes than phantom-like models ($w < -1$), consistent with enhanced late-time decay of gravitational potentials.
    \bigskip
    \item[\textbullet] Taken together, these results from the Gower Street simulations demonstrate that our $w$CDM ISW maps are both accurate and internally consistent, providing reliable predictions for the ISW signal's cosmological dependence in a wide range of models beyond $\Lambda$CDM. 
\end{itemize}

With our methodology, we have established that we can robustly model the 2-point statistics of ISW maps, and our map-based approach also paves the way for higher-order statistics. In subsequent analyses, we expect to use the density and ISW maps for various cross-correlation measurements, including stacking methods and simulation-based inference, searching for optimal ways to detect deviations from the $\Lambda$CDM model via the ISW effect. We also plan to extend this analysis to $w_{0}w_{a}$CDM and perhaps other alternative models, which may offer new insights about dark energy's dynamical properties using exquisite data from surveys like \emph{Euclid}, DESI, Rubin-LSST, and \emph{Roman} in the coming years.

\section*{Acknowledgments}
The authors thank Barbara Mat\'ecsa for assistance with preliminary tests.  
We also thank Krishna Naidoo, whose work formed the foundation of this paper and who provided valuable guidance and assistance throughout our research. We also thank Susan Pyne for constructive comments on the manuscript.

The Large-Scale Structure (LSS) research group at Konkoly Observatory has been supported by a \emph{Lend\"ulet} excellence grant by the Hungarian Academy of Sciences (MTA). This project has received funding from the European Union’s Horizon Europe research and innovation programme under the Marie Skłodowska-Curie grant agreement number 101130774. Funding for this project was also available in part through the Hungarian National Research, Development and Innovation Office (NKFIH, grant OTKA NN147550).

M.G.Y. was supported by the EK\"OP-25 University Excellence Scholarship Program of the Ministry for Culture and Innovation from the source of the National Research, Development and Innovation Fund.

IS acknowledges NASA grant 80NSSC24K1489. The Gower Street simulations were generated under the DiRAC project p153 ‘Likelihood-free inference with the Dark Energy Survey’ (ACSP255/ACSC1) using DiRAC (STFC) HPC facilities (www.
dirac.ac.uk).

We are grateful for the possibility to use HUN-REN Cloud (see \cite{Heder2022}; \hyperlink{https://science-cloud.hu/}{https://science-cloud.hu/}), which helped us achieve the results published in this paper.

\bibliographystyle{aa}
\bibliography{refs}

\end{document}